# Intelligent Approaches to interact with Machines using Hand Gesture Recognition in Natural way: A Survey

Ankit Chaudhary[1], J. L. Raheja[2], Karen Das[3], Sonia Raheja[4]

[1]Computer Vision Research Group, BITS, Pilani, INDIA
`ankitc.bitspilani@gmail.com`
[2] Machine Vision Lab, Digital Systems Group, CEERI,Pilani,INDIA
`jagdish@ceeri.ernet.in`
[3]Tezpur University, Assam, INDIA
`karenkdas@gmail.com`
[4]`soniaraheja@rediffmail.com`

*Abstract.* Hand gestures recognition (HGR) is one of the main areas of research for the engineers, scientists and bioinformatics. HGR is the natural way of Human Machine interaction and today many researchers in the academia and industry are working on different application to make interactions more easy, natural and convenient without wearing any extra device. HGR can be applied from games control to vision enabled robot control, from virtual reality to smart home systems. In this paper we are discussing work done in the area of hand gesture recognition where focus is on the intelligent approaches including soft computing based methods like artificial neural network, fuzzy logic, genetic algorithms etc. The methods in the preprocessing of image for segmentation and hand image construction also taken into study. Most researchers used fingertips for hand detection in appearance based modeling. Finally the comparison of results given by different researchers is also presented.

## 1 Introduction

Natural HGR is one of the very active research areas in the Computer Vision field. It provides the easiness to interact with machines without using any extra device and if the users don't have much technical knowledge about the system, they still will be able to use the system with their normal hands. Gestures communicate the meaning of statement said by the human being. They come naturally with the words to help the receiver to understand the communication. It allows individuals to communicate feelings and thoughts with different emotions with words or without words [1]. This paper is the extended version of our previous work, which we performed to find the current state of the art in hand gesture recognition [51] in context of soft computing. Gesture made by human being can be any but few have a special meaning. Human hand can have movement in any direction and can bend to any angle in all available coordinates. Chinese sign language as shown in figure 1, used hand gestures to represents digits as well as alphabets. Many researchers [4][7][16][39][40] have tried with different instruments and equipment to measure hand movements like gloves, sensors or wires, but in these techniques user have to wear the device which doesn't make sense in practical use. So people thought about a way of contact less gesture recognition that could be considered as a research area in Machine Vision or Computer Vision and which would be as natural as human to human interaction. According to Mitra [6] gesture recognition is a process where user made gesture and receiver recognize it. Using this technique, we can easily interact with machines and can give them particular message according to the environment and application syntax. Even people who can't communicate orally (sick, old or young child), they would also get benefit from this technology. It is possible to make a gesture recognition system for these people. Mobile companies are trying to make handsets which can recognize gesture and could operate from little distance also [2][47]. Here we are focusing on human to machine interaction (HMI), in which machine would be able to recognize the gesture made by human. There are approaches of two types.
   a. Appearance based approaches where hand image is reconstructed using the image properties and extraction.
   b. Model based approaches where different models are used to model image using different models to represent in Computers.

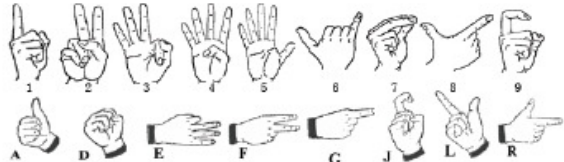
**Fig 1:** Chinese sign language [41]

Here we are dividing approaches based on the method used in it not on how it is treating the image. Many approaches have been developed to interact with machines from glove based [4] to neural networks [3]. Users always like the easy and naturalness of technology in HMI and it was more convenient to interpret visual inputs [7]. As Pickering stated [22] that initially touch based gesture interfaces would be popular but, non-contact gesture recognition technologies would be more attractive finally. Input to a machine using gesture is simple and convenient, but the communication includes many difficulties. "The human hand is a complex deformable object and gesture itself has many characteristics, such as diversities, ambiguities, temporal and spatial differences and human vision itself is an ill-posed problem" [10]. Pickering [22] described a real time gesture based driving system simulator developed at Carnegie Mellon University with the help of General Motors. Many researchers [25][26] [27][28][37][34] have used a color strip or a shirt to detect hand image in the captured image. For a detailed survey of gesture recognition you can see [6][7][23]. Gesture segmentation a part of the gesture recognition process, have been reviewed in [10] and [11] based on color spaces.

Choi [15] brings attention of researchers pointing out an old problem of the incrementing processing time of algorithm's complexity and say "the most important issue in field of the gesture recognition is the simplification of algorithm and the reduction of processing time". He used morphological operation to implement his system using the center points extracted from primitive elements by morphological shape decomposition. Lu [19], Gastaldi [29], Ozer [30] used parallel approach in the design and implementation of their system. Different threads are implemented in such way that they can run in parallel and can compute faster. Shin [61] presented a 3D system HGR system with the application in fruit fly chromosomes based on 2D slices of CT scan images. Lee [17] describes his system which he developed for remote control systems which worked for motion recognition also. He uses 3D systems with two or more cameras to detect command issued by hand. Villani [9] has tried to develop a system for teaching mathematics to the deaf with an easy user interface. Morimoto [43] made interesting virtual system, in which he pushed virtual buttons using fingers in the air and recognized it using 3D sensors.

## 2    Hand Detection Approaches

There are many techniques to detect hand in the acquired image after preprocessing. As shown above, we divide these approaches into two parts.

### 2.1  Appearance Based Approaches

Many researchers have used fingertip detection for the hand image construction

[3][8][12][13][17][21][29][34][37][44][57][62]. As we are also using fingertip detection technique for our research work, this paper devotes great attention to work done by other researches using this technique. Nolker [3] focuses on large number of 3D hand postures in her system called GREFIT. She used finger tips in hands as natural determinant of hand posture to reconstruct the image. In her system she suggests few approaches to locate fingertip in hand.

1. Marked fingertips colored and making histogram
2. Using different templates or images of a prototype

It takes 192x144 size gray scale image to process. Verma [8] extract features from image as fingertip, edges and vectors for 2D modeling. He used harris corner detector to extract fingertips corresponding to corners. Nguyen [12] used gray scale morphology and geometric calculations to relocate fingertip locations using learning based model on 640x480 pixel size frame. Here Author use similar approach to hand detector given by shin [13] to detect both hands based on skin color. To recognize hands Nguyen [12] used skin segmentation technique using Gaussian model. Density function of skin color distribution is as defined.

$$p(c|skin) = \sum_{i=1}^{k} \pi_i \, p_i(c|skin)$$

Where k is the number of components and $\pi_i$ are the weight factors of each component. He used CIELUV color space to represent skin. Interestingly he used palm to finger length ratio to construct the hand image. Zhou [21] worked with 320x240 size 24 bit image frames. Zhou used Markov Random Field to remove noise component in processed image.

Gastaldi [29] find perimeter using Gaussian filters and freeman's algorithm [31] to localize fingertips in that image for 3D detection. Kim [37] tried to recognize gesture in a dark room on black projection for his system. Although the system was vision based but he used florescent white paper to mark finger tips in the captured image, which is not practical for generic purpose as user have to wear white florescent strips. Kim used kalman filter for finding fingertips and their correct positions in a recursive manner. Stefan [5] implemented a system which can detect motion of fingers in the air visually. He made it to recognize the numbers for 0 to 9 for command transfer. Ng [54] developed a system to recognize the 14 predefined gestures in 320x240 pixel sizes in 24 bit color, where hands were moving and the system was able to work with one or both hands. Ng performed a wrist cutting operation on hand images to make both image invariable.

**2.2 Model Based Approaches**

Sawah [34] used histogram for calculating probability for skin color observation. Hu [38] take Gaussian distribution for background pixels marking then he subtracted the pixels from the new image to acquired gesture image. Lee [18] used the same technique to get gesture image.

$$\Delta = | I_n - B |$$

In the modeling of his application of human activity monitoring, Hu [38] applied Genetic Algorithm (GA) to Chromosome pool with $P_{c0}$ and $P_{m0}$ as crossover and mutation rate respectively which he founded using different statistic attributes. Crossover creates new chromosomes while mutation in this case introduces new genes into chromosome. Lee [44] use $YC_bC_r$ skin color model to detect hand region and then he applied distance transform. Tarrataca [47] used RGB and HSI color space model based algorithm for skin detection. Malassiotis [55] developed a system to recognize real time hand gestures in German sign language in 3D using a sensor enabled camera which can find the pattern based on illumination and computes the 3D coordinates of each point on surface. The details about the pattern finding 3D coordinates are given in his other paper [56]. Lien [62] presented a model base system for HGR where the joints in fingers had one DOF (degree of freedom), effective joins had 2 DOF and spherical joints had 3 DOF. So fingers had 4 DOF while thumb had 5. Then he defined local coordinate systems with the origins on the joints (figure 2). This system was interdependent on fingers movement. He used fast fitting method and find the angles of each fingers.

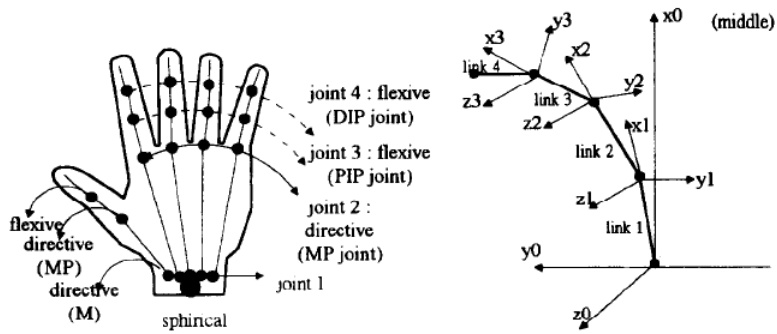

**Fig 2:** (a) hand model (b) local coordinate frames on the joint position for middle finger [62]

## 3  Soft Computing Approaches

Under the umbrella of soft computing principal constituents are Neural Networks, Fuzzy Systems, Machine Learning, Evolutionary Computation, Probabilistic Reasoning, etc. and their hybrid approaches. Here we are focusing on mainly three components:-

a. Artificial Neural Networks
b. Fuzzy Logic
c. Genetic Algorithm

### 3.1  Artificial Neural Network

An Artificial Neural Network (ANN) is made of many highly interconnected processing elements, which are working in together to solve specific problems [35]. ANN can be configured for problems like pattern recognitions or data mining through learning based models. Also ANN has capabilities like adaptive learning, self-organizing and real time operations using special hardware. Nolker [3] used ANN based layer approach to detect fingertips. After obtaining fingertips vectors, it is transformed into finger joint angles to an articulated hand model. For each finger separate network were trained on same feature vectors, having input space 35 dimensional while output dimensional as only 2. Lee [17] used Hidden Markov Model (HMM) for gesture recognition using shape feature. Gesture state is determined after stabilizing the image component as open fingers in consecutive frames. He also used maxima and minima approach like Raheja [14] for construction the hand image and FSM like Verma [8] for gesture finalization.

   Wang [32] proposed an optical flow based powerful approach for human action recognition using learning models. It labels hidden parts in image also. This mas-margin based algorithm can be applied to gesture recognition. Kim [37] in his system used learning model for dynamic gestures recognition. Ng [52] used HMM and RNNs for gesture classification from the collected vectors of hand pose frames. Outputs of both classifiers were combined to get better result and it was input to the developed GUI. They used Fourier descriptors to represent the boundary of extracted binary hand and trained Radial Basis Function consisted of 56 input nodes, 38 hidden layer and five output nodes. The activation function of the j$^{th}$ hidden node was given by

$$\varphi_j(\mathbf{x}) = \exp\left(-\frac{\|\mathbf{x} - \mathbf{c}_j\|^2}{2\sigma_j^2}\right)$$

Where x in input vector, $c_j$ is the center and $\sigma_j$ is the spread of $\varphi_j(x)$. Just [60] has presented a comparative study of HMM and IOHMM HGR techniques on the two openly accessible databases, and came with the conclusion that HMM is a better choice for the HGR modeling. Stergiopoulou [59] used unsupervised Self-Growing and Self-Organized Neural Gas network for 31 pre specified gesture. Although he made several assumptions like arm should be vertical, and user is using only his right hand, while this system have a problem when left handed users is showing the gesture. The raised fingers detection in the hand is done by finding the fingertip neuron, which would be followed by the other neurons chain as shown in figure 3. The center of palm could be calculated by gravity method from neuron only in palm area and distance from the fingertips to palm center is calculated, but the main problem in gesture recognition is that only raised fingers would be counted in presented algorithm (figure 3) and gesture would be recognized accordingly. Then he applied a likely hood classification to get the gestures which are predefined based on raised fingers.

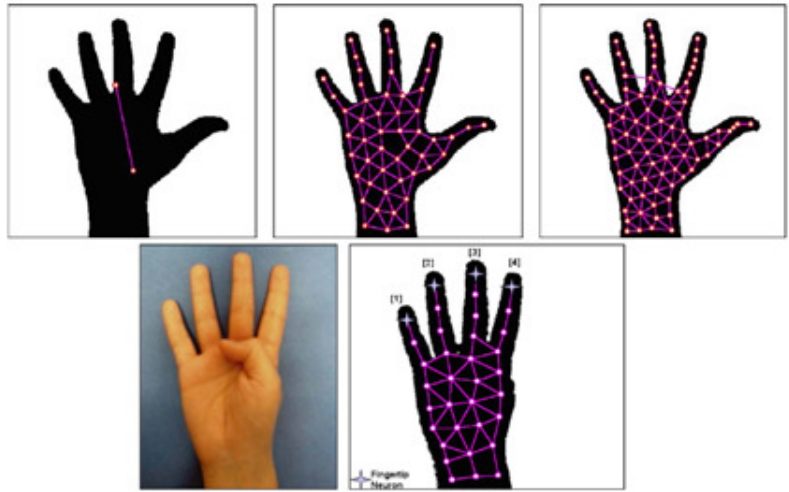

**Fig 3:** SGONG network working (a) start with two points (b) growing stage with 45 neurons (c) output with 83 neurons (d) hand gesture (e) only raised fingers would be counted [59]

### 3.2 Fuzzy Logic Based Approaches

A Professor from UCB USA, Lotfi Zadeh presented fuzzy logic in an innovative way. His view was that for processing precise and accurate information is not necessary, we can perform it with imprecise data also. It is near to natural thinking. As described in [35] "Fuzzy logic is a multivalued logic that allows intermediate values to be defined between conventional evaluations". Verma [8] used c-mean fuzzy clustering based finite state machines (FSM) to recognize hand gestures. Formula for centroid calculation of fuzzy c-means clusters is that centroid would be mean of all points weighted by their degree of belongings to the cluster center. For each point x, a coefficient giving the degree in the $k^{th}$ cluster $U_k(x)$ [24]. Here $x_k = k^{th}$ trajectory point, so

$$\text{center}_k = \frac{\sum_x u_x(x)^m x}{\sum_x u_x(x)^m}$$

In second phase these cluster maps onto FSM states and final state show gesture recognition, although Verma [8] didn't implement it. Schlomer [45] used k-mean algorithm on clusters, then he applied HMM and Bayes-classifier on vector data. Trivino [36] tried to make a more descriptive system which can convert human gesture positions into a linguistic description using fuzzy logic. He related it to Natural Language Processing (NLP). He used sensors and took only few positions in sitting and standing, into consideration.

### 3.3 Genetic algorithm based approaches

Genetic Algorithm comes from biology but it is very influential on computational sciences in optimization. This method is very effective to get optimal or sub optimal solutions of problems as it have only few constraints [35]. It uses generate and test mechanism over a set of probable solutions (called as population in GA) and bring optimal acceptable solution. It executes its three basic operations (Reproduction, Crossover and Mutation) iteratively on population. Sawah [34] has focused on a very generic scenario where he used generic non-restricted environment, generic not-specific application for gesture recognition using genetic programming. He used crossover for noise removal in gesture recognition, while Dynamic Bayesian Network (DBN) for gesture segmentation and gesture recognition with the fuzzification. Hu [38] applied Genetic Algorithm on his system which make 2D parametric model with human silhouette in his application of Human Activity Monitoring. The best point about GA is that it work parallel on different points for faster computation.

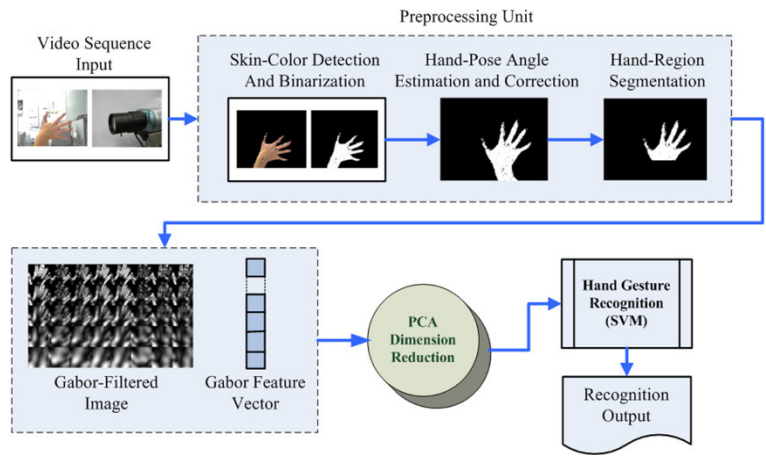

**Fig 4:** Hand Gesture Recognition process form video [52]

### 3.4 Other Approaches

Raheja [14] proposes a new methodology for real time robot control using Principal Component Analysis (PCA) for gesture extraction and pattern recognition with saved images in database in 60x80 image pixels formats. He used syntax of few gestures and decides corresponding actions of robot. He claims that PCA method is very faster than neural network based methods which require training database and more computation power. Huang [52] used PCA for dimensionality reduction and SVM for gesture classification in using skin color model switching for varying illumination environment. In Huang approach (Figure 4) image sequences was sent for skin-color detection, hand pose angle estimation and hand region segmentation. Then it divide resultant image into 40, 20x20 pixels size and run Gabor filter. Morimoto [43] also used PCA and maxima methods. Gastaldi [29] used PCA to compress five image sequences into one and get eigen vectors and eigen values for each gesture. He used statistical HMM model for gesture recognition. Zaki [58] used PCA where hand representation is transformed from the image coordinates to eigen vector space. After vector rotation the largest eigen vector was aligned with the mid of data as shown in figure 5. He used three HMMs for every sign, one for each feature PCA Sequence, Kurt Pos Sequence and MCC Sequence found.

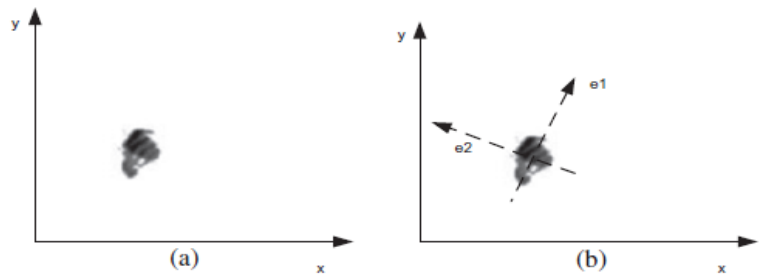

**Fig 5:** Hand (a) coordinates and (b) eigen vectors [58]

Shin [33] shows gesture extraction and recognition using entropy analysis and low level image processing functions. Lee [18] also used entropy to get color information. He used PIM to quantify the entropy of image using the following equation.

$$PIM = \sum_{i=0}^{L-1} h(i) - Max_j\, h(i)$$

Where $h(i)$ is the $i^{th}$ histogram value of each image or block. To acquire PIM value, subtracting all pixels in each block from maximum frequency in histogram model.

Lu [19] implemented system for 3D gesture recognition where he fused different positions of gesture using coordinate transformations and then use stored prespecified gestures for gesture recognition. Stefan [5] has used Dynamic Space-Time Warping (DSTW) [42] to recognize a set of gestures. This technique doesn't require hands to be correctly identified at each frame. Zou [46] used Deterministic Finite State Machine (DFSM) to detect hand motion and then apply rule based techniques for gesture recognition. He defines gesture into two category based on motion linear and arc shaped gestures. Tarrataca [47] used convex hull method based clustering algorithm Graham's Scan [48] for posture recognition. Chang [49] used a feature alignment approach based on curvature scale space to recognize hand posture.

## 4 Implementation Tools

Mostly researchers who used image processing used MATLAB® with image processing toolbox while few used C++ also. Lu [19], Lee [44] and Zou [46] used C++ for implementation on Windows XP® where Lu [19] and Lee [44] he used Microsoft® Foundation Classes (MFC) to build user interface and control. Intel OpenCV Library is also popular, used with MATLAB® to implement systems [53]. Stergiopoulou [59] used Delphi to implement HGR system using SGONG network.

## 5     Accuracy

GREFIT [3] system was able to detect finger tips even when it was in front of palm, it reconstruct the 3D image of hand that was visually comparable. Nguyen [12] claimed results 90-95% accurate for open fingers that is quite acceptable while for closed finger it was 10-20% only. As shown in figure 6 closed or bended finger are coming in front of palm, so skin color detection would not make any difference in palm or finger. According to him image quality and morphology operator was the main reason for low detection. Raheja [14] claims about 90% accuracy in the result, if the lighting conditions are good. Hu [38] used six different parameters to control the performance of system, if he found much noise there, he could control it using two parameters called as α and β respectively. Stergiopoulou [59] claims about 90.45% accuracy, through hidden finger was not detected in his approach. Morimoto [43] claimed for his system near 91% accuracy after he applied normalization. Ng and Ranganath [54] showed 91.9% correct results with the combination of HMM and RNNs in their approach. Huang [52] claimed 93.7% recognition results using Gabor filters. Lee [18] showed results for six kinds of gesture with recognition rate of more than 95% but it recognized bended finger as bended, no matter the degree of banding. Stefan [5] achieved 96% accuracy over 300 tests. He also stated that parallel image processing and pattern matching operations were not real time compatible in MATLAB®, it could be faster if implemented in C++. For running gestures like Bye and showing a direction, Suk [53] claims 80.77 % results.

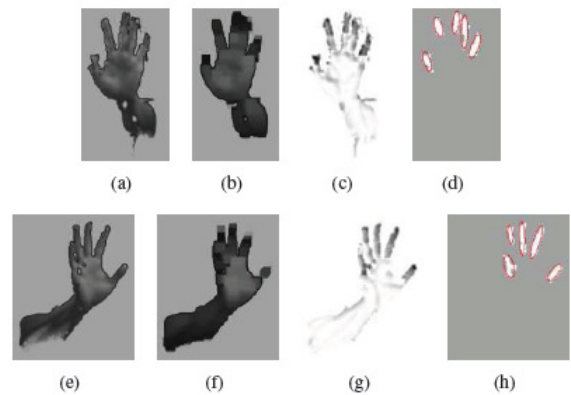

**Fig 6:** Result of finger extraction using grayscale morphology operators and object analysis [12] which work for bended finger also, but with a lower accuracy 10-20%.

## 6     Conclusions

Different applications of hand gesture recognition have been implemented in different domains from simply game inputs to critical applications. Hand gesture recognitions is the natural to interact with vision enabled computers and other machines. This paper primarily focused on the study of work done in the area of natural hand gesture recognition using Computer Vision Techniques.  We did survey based on intelligent approaches mainly in the context of soft computing. Approaches using Artificial Neural Network, Fuzzy Logic, Genetic Algorithm and other well performed intelligent techniques have been discussed and compared. In appearance based approach, main focus was on fingertip detection as it was used by mostly researchers. Soft computing provides a way to define things which are not certain but with an approximation that can be make sure using learning models and training data. So soft computing is very effective in getting the results where the exact positions of hand or fingers are not possible. In the future we will work in the area of individual finger position bending detection and movements, as work done in this area are very few. Mostly researchers worked with full hand position detection or the fingertip position to write virtual words.

## 7      Acknowledgements

Authors would like to thank Director, CEERI Pilani for providing research facilities and to The Chief, Information Processing Center, BITS Pilani and Head of Department, Computer Science, BITS Pilani for their support and encouragement.

## References

1. Kendon A. Gesture: Visible Action as Utterance, Cambridge University Press. UK, 2004.
2. Kroeker K.L., Alternate interface technologies emerge, Communications of the ACM, Vol. 53, No. 2, Feb 2010, pp. 13-15.
3. Nolker C., Ritter H., Visual Recognition of Continuous Hand Postures, IEEE Transactions on neural networks, Vol 13, No. 4, July 2002, pp. 983-994.
4. Sturman D., Zeltzer D., A survey of glove-based input, IEEE Transactions on Computer Graphics and Applications, Vol. 14, No. 1, Jan. 1994, pp. 30-39.
5. Stefan A., Athitsos V., Alon J., Sclaroff S., Translation and scale invariant gesture recognition in complex scenes, Proceedings of 1st international conference on pervasive technologies related to assistive environments, Greece, July 2008.
6. Mitra S., Acharya T., Gesture recognition: a survey, IEEE transactions on systems, man, and cybernetics-part C: applications and review. Vol 37. No 3. May 2007. pp. 2127-2130.
7. Pavlovic V.I., Sharma R., Huang T.S. Visual interpretation of hand gestures for human- computer interaction: A review. IEEE Transactions on Pattern Analysis and Machine Intelligence, Vol 19, July 1997, pp. 677-695.
8. Verma R., Dev A., Vision based Hand Gesture Recognition Using finite State Machines and Fuzzy Logic, International Conference on Ultra-Modern Telecommunications & Workshops, 12-14 Oct, 2009, pp. 1-6.
9. Villani N.A., Heisler J., Arns L., Two gesture recognition systems for immersive math education of the deaf, Proceedings of the first International conference on immersive telecommunications, Bussolengo, Verona, Italy, Oct, 2007.
10. Xu Z., Zhu H., Vision-based detection of dynamic gesture, International Conference on Test and Measurement, 5-6 Dec, 2009, pp.223-226.
11. Mahmoudi F., Parviz M., Visual Hand Tracking algorithms, Geometric Modeling and Imaging- New Trends, 16-18 Aug, 2006, pp. 228-232.
12. Nguyen D.D., Pham T.C., Jeon J.W., Fingertip Detection with Morphology and Geometric Calculation, IEEE/RSJ International Conference on Intelligent Robots and Systems, St. Louis ,USA, Oct 11-15, 2009, pp. 1460-1465.
13. Shin M. C., Tsap L. V., and Goldgof D. B., Gesture recognition using bezier curves for visualization navigation from registered 3-d data, Pattern Recognition, Vol. 37, Issue 5, May 2004, pp.1011–1024.
14. Raheja J.L., Shyam R,. Kumar U., Prasad P.B., Real-Time Robotic Hand Control using Hand Gesture, 2nd international conference on Machine Learning and Computing, 9-11 Feb, 2010, Bangalore, India, pp. 12-16.
15. Choi, J., Ko N., Ko D., Morphological Gesture Recognition Algorithm, Proceeding of IEEE region 10th international conference on Electrical and Electroic Technology, Coimbra, Portugal, 19-22 Aug,2001, pp.291-296.

International Journal of Computer Science & Engineering Survey (IJCSES) Vol.2, No.1, Feb 201116. Cho O.Y., and et al. A hand gestue recognition system for interactive virtual environment, IEEK, 36-s(4), 1999, pp. 70-82.
17. Lee D. and Park Y., Vision-Based Remote Control System by Motion Detection and Open Finger Counting, IEEE Transactions on Consumer Electronics, Vol. 55, issue 4, Nov 2009, pp. 2308-2313.
18. Lee J., and et al., Hand region extraction and gesture recognition from video stream with complex background through entropy analysis, Proceedings of 26th annual international conference of the IEEE EMBS, San Francisco, CA, USA, 1-5 Sep, 2004, pp. 1513-1516.
19. Lu G. and et al. , Dynamic hand gesture tracking and recognition for real time immersive virtual object manipulation, International conference on cyber worlds, 7-11 Sep, 2009, pp. 29-35.
20. Kota S.R. and et al., Principal Component analysis for Gesture Recognition Using SystemC, International Conference on Advances in recent technologies in Communication and Computing, 2009.
21. Zhou H., Ruan Q., A Real-time Gesture Recognition Algorithm on Video Surveillance, 8th international conference on Signal Processing, 2006.
22. Pickering C.A.,The search for a safer driver interface: a review of gesture recognition Human Machine Interface, IEE Computing and Control Engineering, 2005, pp. 34-40.
23. Ong S.C.W., Ranganath S., Automatic Sign Language Analysis: A Survey and the Future beyond Lexical Meaning, IEEE Transactions on pattern analysis and machine intelligence, Vol.27, No. 6, June 2005.
24. Wikipedia.org,http://en.wikipedia.org/wiki/Cluster_analysis#Fuzzy_c-means_clustering.
25. Do J. and et al, Advanced soft remote control system using hand gestures, MICAI (Advances in Artificial Intelligence) 2006, LNAI, vol. 4293, 2006, pp. 745-755.
26. Premaratne P. and Nguyen Q., Consumer electronics control system based on hand gesture moment invariants, IET Computer Vision, vol. 1,no. 1, Mar. 2007, pp. 35-41.
27. Kohler M., Vision based remote control in intelligent home environments, 3D Image Analysis and Synthesis, 1996, pp. 147-154.
28. Bretzner L., Laptev I., Lindeberg T., Lenman S. and Sundblad Y., A Prototype system for computer vision based human computer interaction, Technical report ISRN KTH/NA/P-01/09-SE, 2001.
29. Gastaldi G. and et al., a man-machine communication system based on the visual analysis of dynamic gestures, International conference on image processing, Genoa, Italy, 11-14 Sep, 2005, pp. 397-400.
30. Ozer I. B., Lu T., Wolf W. Design of a Real Time Gesture Recognition System : High Performance through algorithms and software. IEEE Signal Processing Magazine. May, 2005, pp 57-64.
31. Freeman H., on the encoding of arbitrary geometric configurations, IRE Transactions on Electronic Computers, EC-10:260–268, July 1985.
32. Wang Y., Mori G., Max-Margin Hidden conditional random fields for human action recognition, IEEE conference on Computer vision and pattern recognition, Miami, Florida, USA, 20-25 June, 2009, pp. 872-879.
33. Shin j., and et al., Hand region extraction and gesture recognition using entropy analysis, International journal of Computer science and network security, Vo. 6, issue 2A, Feb 2006.
34. Sawah A.E., and et al., a framework for 3D hand tracking and gesture recognition using elements of genetic programming, 4th Canadian conference on Computer and robot vision, Montreal, Canada, 28-30 May, 2007, pp. 495-502.
35. Sivanandam S.N., Deepa S. N., Principles of soft computing, Wiley India Edition, New Delhi, 2007.
36. Trivino G., Bailador G., Linguistic description of human body posture using fuzzy logic and several levels of abstraction, IEEE conference on Computational Intelligence for measurement systems and applications, Ostuni, Italy 27-29 Jun, 2007, pp. 105-109.
37. Kim H., Fellner D. W., Interaction with hand gesture for a back-projection wall, Proceedings of Computer Graphics International, 19 Jun, 2004, pp. 395-402.
38. Hu C., Yu Q., Li Y., Ma S., Extraction of Parametric Human model for posture recognition using Genetic Algorithm, 4th IEEE international conference on automatic face and gesture recognition, Grenoble, France, 28-30 Mar, 2000, pp. 518-523.
39. Huang T.S., and Pavlovic V.I., Hand gesture modeling, analysis and synthesis, Proceedings of international workshop on automatic face and gesture recognition, 1995, pp.73-79.
40. Quek F.K.H., Toward a vision-based hand gesture interface, proceedings of the virtual reality system technology conference, 1994, pp.17-29.
131

41. Zhang J., Lin H., and Zhao M. A Fast Algorithm for Hand Gesture Recongnition using Relief. Sixth International Conference on Fuzzy Systems and Knowledge Discovery. Tinajin, China .14-16 Aug, 2009. pp 8-12.
42. Alon J. and et al. Simultaneous localization and recognition of dynamic hand gestures, International IEEE motion workshop, 2005, pp. 254-260.
43. Morimoto K. and et al, Statistical segmentation and recognition of fingertip trajectories for a gesture interface, Proceedings of the 9th international conference on Multimodal interfaces, Nagoya, Aichi, Japan, 12-15 Nov, 2007, pp. 54-57.
44. Lee B., Chun J., Manipulation of virtual objects in marker-less AR system by fingertip tracking and hand gesture recognition, Proceedings of 2nd international conference on interaction science: Information Technology, Culture and Human, Seoul, Korea, 2009, pp. 1110-1115.
45. Schlomer T. and et al., Gesture recognition with a Wii Controller, Proceedings of the 2nd international conference and embedded interaction, Bonn, Germany, 18-20 Feb, 2008, pp 11-14.
46. Zou S., Xiao H., Wan H., Zhou X., Vision based hand interaction and its application in pervasive games, Proceedings of the 8th international conference on virtual reality continuum and its applications in industry, Yokohama, Japan, 2009, pp. 157-162.
47. Tarrataca L., Santos A.C. and Cardoso J.M.P., The current feasibility of gesture recognition for a smartphone using J2ME, Proceedings of the 2009 ACM symposium on applied computing, pp.1642-1649.
48. Graham R., An efficient algorithm for determining the convex hull of a finite planar set, information processing letters, 1972, Vol. 13, pp. 21-27.
49. Chang C., Liu C., Tai W., Feature alignment approach for hand posture recognition based on curvature scale space, Neurocomputing, Neurocomputing for Vision Research; Advances in Blind Signal Processing, Jun 2008, Vol. 71, Issues 10-12, pp. 1947-1953.
50. Li H. and Greenspan M., Model-based segmentation and recognition of dynamic gestures in continuous video streams, *Pattern Recognition*, doi:10.1016/j.patcog.2010.12.014.
51. Chaudhary A. and et al., A Survey on Hand Gesture Recognition in context of Soft Computing, Lecture Notes in Computer Science (CCIS, SPRINGER VERLAG), CCSIT 2011, Part III, CCIS 133, pp. 46-55.
52. Huang D., Hu W., Chang S., Gabor filter-based hand-pose angle estimation for hand gesture recognition under varying illumination, Expert Systems with Applications, DOI: 10.1016/j.eswa.2010.11.016.
53. Suk H., Sin B., Lee S., Hand gesture recognition based on dynamic Bayesian network framework, Pattern Recognition, Vol. 43, Issue 9, Sep 2010, pp 3059-3072.
54. Ng C.W., Ranganath S., Real-time gesture recognition system and application, Image and Vision Computing, Vol. 20, Issues 13-14, Dec 2002, pp. 993-1007.
55. Malassiotis S., Strintzis M.G., Real-time hand posture recognition using range data, Image and Vision Computing, Vol. 26, Issue 7, 2 Jul 2008, pp. 1027-1037.
56. Tsalakanidou F.,Forster F., Malassiotis S., Strintzis M.G., Real-time acquisition of depth and color images using structured light and its application to 3d face recognition, Real-Time Imaging, Special Issue on Multi-Dimensional Image Processing. Vol. 11, Issue 5–6, 2005.
57. Chaudhary A. and Raheja J.L., ABHIVYAKTI: A Vision Based Intelligent Pervasive System for Elder and Sick Persons, the 3rd IEEE International Conference on Machine Vision, Hong Kong, 28-30 Dec, 2010, pp. 361-364.
58. Zaki M. M., Shaheen S. I., Sign language recognition using a combination of new vision based features, Pattern Recognition Letters, Vol. 32, Issue 4, 1 Mar 2011, pp. 572-577.
59. Stergiopoulou E., Papamarkos N., Hand gesture recognition using a neural network shape fitting technique, Engineering Applications of Artificial Intelligence, Vol. 22, Issue 8, Dec 2009, pp. 1141-1158.
60. Just A., Marcel S., A comparative study of two state-of-the-art sequence processing techniques for hand gesture recognition, Computer Vision and Image Understanding, Vol. 113, Issue 4, Apr 2009, pp. 532-543.
61. Shin M. C., Tsap L. V., Goldgof D. B., Gesture recognition using Bezier curves for visualization navigation from registered 3-D data, Pattern Recognition, Vol. 37, Issue 5, May 2004, pp 1011-1024.
62. Lien C.C., Huang C., Model-based articulated hand motion tracking for gesture recognition, Image and Vision Computing, Vol. 16, Issue 2, 20 Feb 1998, pp. 121-134.

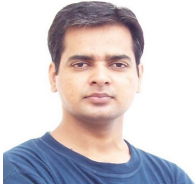
Ankit Chaudhary received his Master of Engineering degree in Computer Science & Engineering from Birla Institute of Technology & Science, Pilani, INDIA and currently working toward his Ph.D. in Computer Vision, from BITS Pilani, INDIA. His areas of research interest are Computer Vision, Digital Image Processing, Artificial Intelligence and Mathematical Computations.

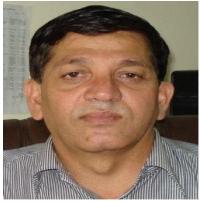
Dr. J.L.Raheja Received his M.Tech from Indian Institute of Technology, Kharagpur, INDIA and Ph.D. from Technical University of Munich, GERMANY. Currently he is senior scientist in Digital Systems Group at Central Electronics Engineering Research Institute (CEERI), Pilani, INDIA.  His areas of research interest are Digital Image Processing, Embedded Systems and Human Computer Interface.

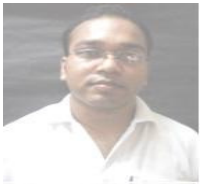
Karen Das received his M.Sc. in Electronics Science from Guwahati University, INDIA and currently pursuing M.Tech, in Electronics Engineering from Tezpur University, Assam. INDIA. His areas of research interest are Digital Image Processing, Artificial Intelligence, VLSI systems, and Digital Communication.